\documentclass{article}

\usepackage[utf8]{inputenc}
\usepackage{amsmath, amsfonts, dsfont}
\usepackage{graphicx}

\usepackage{longtable, booktabs}

\usepackage[]{natbib}

\usepackage{geometry}
\title{Active extension portfolio optimization with non-convex risk measures
using metaheuristics}

\author{
Ronald Hochreiter\and
Christoph Waldhauser\and
}

\begin{document}

\maketitle

\begin{abstract}
We consider the optimization of active extension portfolios. For this
purpose, the optimization problem is rewritten as a stochastic
programming model and solved using a clever multi-start local search
heuristic, which turns out to provide stable solutions. The heuristic
solutions are compared to optimization results of convex optimization
solvers where applicable. Furthermore, the approach is applied to solve
problems with non-convex risk measures, most notably to minimize
Value-at-Risk. Numerical results using data from both the Dow Jones
Industrial Average as well as the DAX 30 are shown.
\end{abstract}

\section{Introduction}\label{introduction}

In this paper we consider the optimization of active extension
portfolios, which are also known as 1x0/x0 (most commonly 130/30)
portfolios, see e.g. \citet{lo2008130}, \citet{gastineau2008short}, and
\citet{thomas2007alpha}. The idea is to extend long-only portfolios to
contain a certain additional percentage $x$ of the investors budget on
the long-side of the portfolio and additionally short $-x$ percent of
the assets in the portfolio. Classical approaches to create active
extension portfolios are often based on sorting returns and applying a
momentum approach. Historically well-performing assets are used for the
additional long part and bad-performing assets for the short part.
Commonly, no real optimization is conducted. The problem is that this
methodology does not fit well into the classical Modern Portfolio Theory
optimization framework, i.e.~the Markowitz approach to calculate
risk-optimal financial portfolios as shown by
\citet{markowitz1952portfolio}. This approach is defined by calculating
a risk-optimal portfolio $x$ given a set of $a$ financial assets for
which a vector of expected returns $M$ and a co-variance matrix
$\mathbb{C}$ exists. Further constraints $\mathcal{X}$ may be added
(e.g.~long and short restrictions, \ldots), i.e.

\[\begin{aligned}
\begin{array}{ll}
\text{minimize } x & x \; \mathbb{C} \; x^T \\
\text{subject to} & x \times M \geq \mu \\
& x \in \mathcal{X}. \\
\end{array}\end{aligned}\]

The issue with this approach is that uncertainty is implicitly modeled,
only the first and second moments of the loss distribution are used.
This is especially problematic in times of some financial crisis.
Furthermore, from an financial point of view the Variance is probably
not the most useful risk measure, because is is penalizing the upside.
From an optimization point of view, the quadratic programming framework
is too rigid to implement additional extensions. The optimization of
expected shortfall objectives and constraints cannot easily be put on
top of this rather specific base model. More importantly, the quadratic
programming objective does not allow for a simple active extension
constraint, because such an approach is based on the solution structure
of linear programs. For this reason, we will apply the technology of
stochastic programming, see \citet{RuszczynskiShapiro2003},
\citet{WallaceZiemba2005}, and \citet{KingWallace2013} for more details
about this technique.

While there are convex optimization reformulations for many important
risk measures (e.g.~the Mean Absolute Deviation as proposed by
\citet{konno1991mean} or the LP-based Conditional Value-at-Risk
(Expected Shortfall) approach by \citet{rockafellar2000optimization} and
\citet{rockafellar2002conditional} respectively), problems arise when an
investor needs to integrate non-convex risk measures like the
Value-at-Risk \citep{jorion1997value}. For this purpose metaheuristics
can prove to be useful. Many metaheuristics have been shown to solve
portfolio optimization problems with a varying degree of success. A good
overview can be found in the three volumes on Natural Computing in
Computational Finance, see \citet{ncfin2008}, \citet{ncfin2009}, and
\citet{ncfin2010}. Most of the presented metaheuristics are complex and
don't scale well given the fact that the underlying optimization problem
is rather simple from a heuristic optimization point of view. Therefore
we propose a simple, yet powerful multi-start local search heuristic,
which integrates structural information of the portfolio optimization
process into its heuristic framework.

This paper is organized as follows. Section
\ref{scenario-based-portfolio-optimization} presents a short overview of
scenario-based portfolio optimization, Section
\ref{multi-start-local-search-heuristics} describes the proposed
heuristic, while Section \ref{numerical-results} provides numerical
results using data from both the Dow Jones Industrial Average as well as
the DAX 30 index. Section \ref{conclusion} concludes the paper.

\section{Scenario-based Portfolio
Optimization}\label{scenario-based-portfolio-optimization}

Stochastic programming is well suited to model optimization problems
under uncertainty, because of its inherent feature to split a model into
an objective and a subjective part explicitly within the optimization
modeling process. In terms of financial portfolio optimization, the
constraint set $\mathcal{X}$ contains e.g.~regulatory and organizational
constraints, i.e.~the objective part. The subjective views on the
underlying uncertainty is expressed by a scenario set $S$, which in this
specific application contains a discrete asset return probability
(uncertainty) model, i.e.~a set of different probable returns for each
asset. Using this scenario set and a heuristic approach, any risk
measure (VaR, Omega, \ldots) can be integrated, because the evaluation
of the respective loss distribution $\ell_x$ for some portfolio $x$,
i.e. $\ell_x = \big < x, S \big >$ can be used to evaluate with any
functional -- independent of its underlying mathematical structure.

An investor usually faces a bi-criteria optimization problem, i.e.~she
wants to maximize the expected return while also minimizing the risk.
The meta-model for this scenario-based stochastic portfolio optimization
problem is thus given by:

\[\begin{aligned}
\begin{array}{lll}
\text{maximize } x & \text{Return}(\ell_x) \\
\text{minimize } x & \text{Risk}(\ell_x) \\
\text{subject to} & x \in \mathcal{X} \\
\end{array}\end{aligned}\]

While the multi-criteria optimization problem is interesting from a
research point of view, in most practical applications we will
reformulate the above optimization problem to a single objective model,
where the risk is minimized in the objective and the return is
controlled via a constraint and a given minimum acceptable lower bound
$\mu$ on the expected return, i.e.

\[\begin{aligned}
\begin{array}{lll}
\text{minimize } x & \text{Risk}(\ell_x) \\
\text{subject to} & \text{Return}(\ell_x) \geq \mu \\
 & x \in \mathcal{X} \\
\end{array}\end{aligned}\]

\section{Multi-start local search
heuristics}\label{multi-start-local-search-heuristics}

We implemented a simple but powerful multi-start local search heuristic
to solve the active extension portfolio optimization problem. It
contains of three parts:

\begin{enumerate}
\def\labelenumi{\arabic{enumi}.}
\item
  Sample a number $n_1$ of random portfolios using a special sampling
  algorithm.
\item
  Improve the best $n_2$ random portfolios using an iterative
  $\varepsilon$-improvement procedure.
\item
  Pick the improvement with the best objective (risk-ratio).
\end{enumerate}

Finally, if the $n_2$ portfolios differ, then the mean of these
portfolios is taken and the iterative $\varepsilon$-improvement is
applied to this resulting portfolio again.

\subsection{Portfolio Sampling}\label{portfolio-sampling}

Drawing random numbers and creating portfolios out of these numbers can
be tricky. This is especially true when one needs good starting points
because of the danger to get stuck in local optima based on the
underlying heuristic technique. This is valid in our case for the
subsequent portfolio $\varepsilon$-improvement. Hence, the general
structure of optimal portfolios needs to be taken into consideration
before sampling portfolios. The main structure is that many real-world
assets in risk-optimal portfolios are not chosen at all, i.e.~only a
small subset is selected. The applied sampling method in this paper
requires the following parameters:

\begin{itemize}
\item
  The amount of long (default: 0.3) and short (default: 0.1) assets in
  percent.
\item
  The upper and lower bound on each asset (default: 0.5 long and $-0.1$
  short).
\item
  The sum of the long (default: 1.30) and short (default: $-0.3$) part
  of the portfolio.
\end{itemize}

Randomly sampled portfolios with these default values exhibit the
well-known sparsity of real-life risk-optimal (active extension)
portfolios.

\subsection{Iterated portfolio
$\varepsilon$-improvement}\label{iterated-portfolio-varepsilon-improvement}

The $\varepsilon$-improvement is a multi-start local search heuristic
based on the $n_2$ best sampled portfolios from the above described
sampling procedure. Each asset is modified by $\pm \varepsilon$,
i.e.~out of the initial portfolio, $a \times 2$ new portfolios are
created. Thereby the lower and upper bounds are easily satisfiable. The
resulting portfolios are normalized to the given sum of both the long
and the short side. Finally, the (locally) best improvement is chosen to
be the next portfolio until no local improvement can be accomplished.

Depending on the structure of the underlying scenario set as well as the
given constraints, a different set of $\varepsilon$ might be applicable.
Simple solutions can be computed with e.g.
$\varepsilon = (0.05, 0.01, 0.001)$, i.e.~the local search will be
repeated three times with a smaller $\varepsilon$ in each run. A broader
set of $\varepsilon$ can be used too.

\section{Numerical results}\label{numerical-results}

We have tested the algorithm with assets of two mayor financial stock
indices -- both the Dow Jones Industrial Average DJIA (containing the
assets AA, AXP, BA, BAC, CAT, CSCO, CVX, DD, DIS, GE, HD, HPQ, IBM,
INTC, JNJ, JPM, KO, MCD, MMM, MRK, MSFT, PFE, PG, T, TRV, UNH, UTX, VZ,
WMT, XOM) with weekly returns from the year 2012 as well as the DAX 30
(containing the assets ADS.DE, ALV.DE, BAS.DE, BAYN.DE, BEI.DE, BMW.DE,
CBK.DE, CON.DE, DAI.DE, DBK.DE, DB1.DE, LHA.DE, DPW.DE, DTE.DE, EOAN.DE,
FRE.DE, FME.DE, HEI.DE, HEN3.DE, IFX.DE, SDF.DE, LXS.DE, LIN.DE, MRK.DE,
MUV2.DE, RWE.DE, SAP.DE, SIE.DE, TKA.DE, VOW3.DE) with weekly returns
from the year 2013. The above mentioned ticker symbols are taken from
Yahoo! Finance, which also served as the data source.

The algorithm was implemented using R \citep{R2013}. The optimization of
the convex linear programs has been conducted with the GNU Linear
Programming Kit 4.54. The convex quadratic optimization problems have
been solved with the \texttt{quadprog} R package, which implements the
optimization algorithm proposed by \citet{goldfarb1983numerically}.

\subsection{Long-only Markowitz
Portfolios}\label{long-only-markowitz-portfolios}

\begin{figure}
\begin{center}
\includegraphics[scale=0.5]{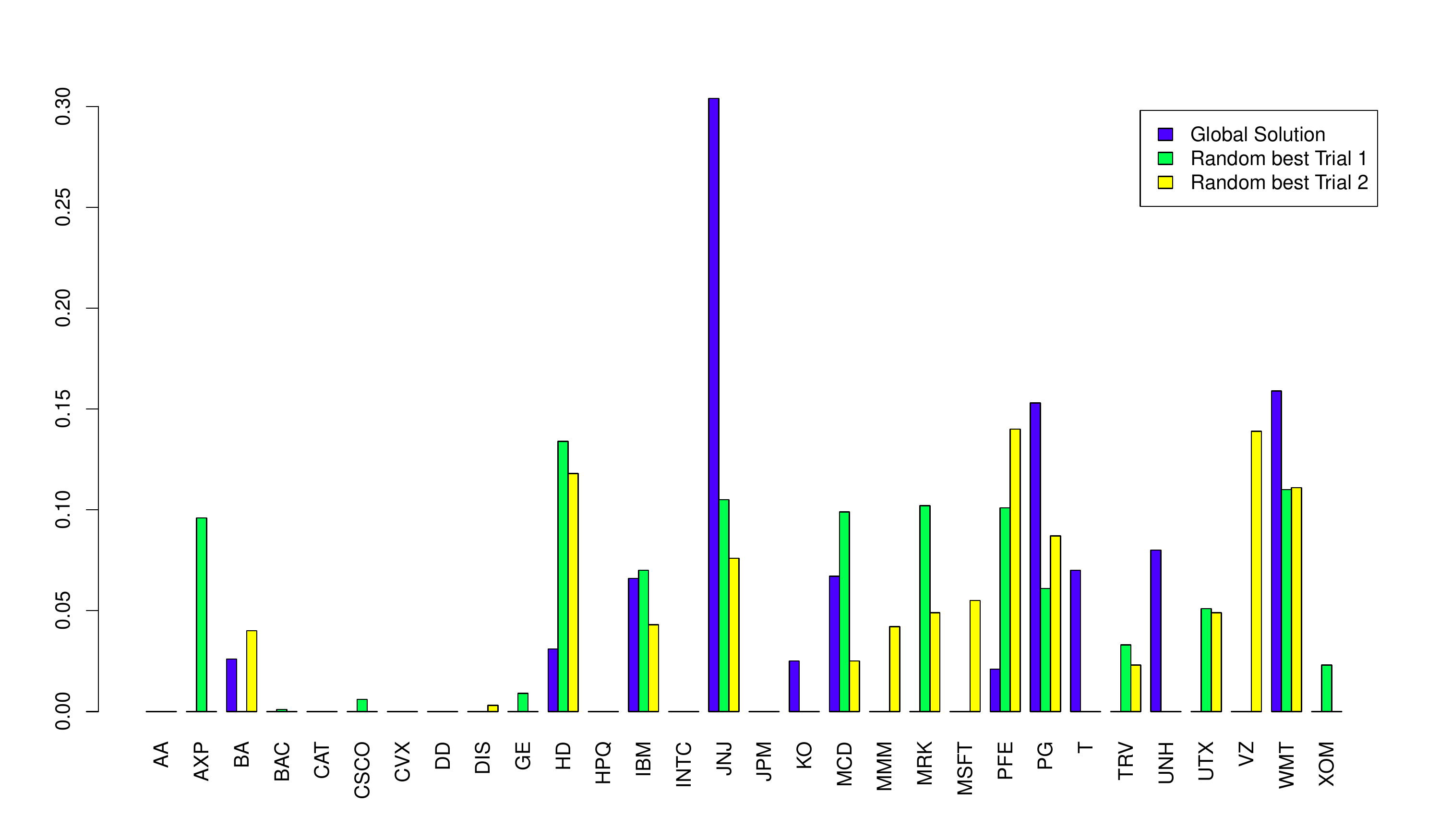}
\caption{Long-only Markowitz-optimal portfolio and sampled portfolios.}
\label{fig:ex1a}
\end{center}
\end{figure}

\begin{figure}
\begin{center}
\includegraphics[scale=0.5]{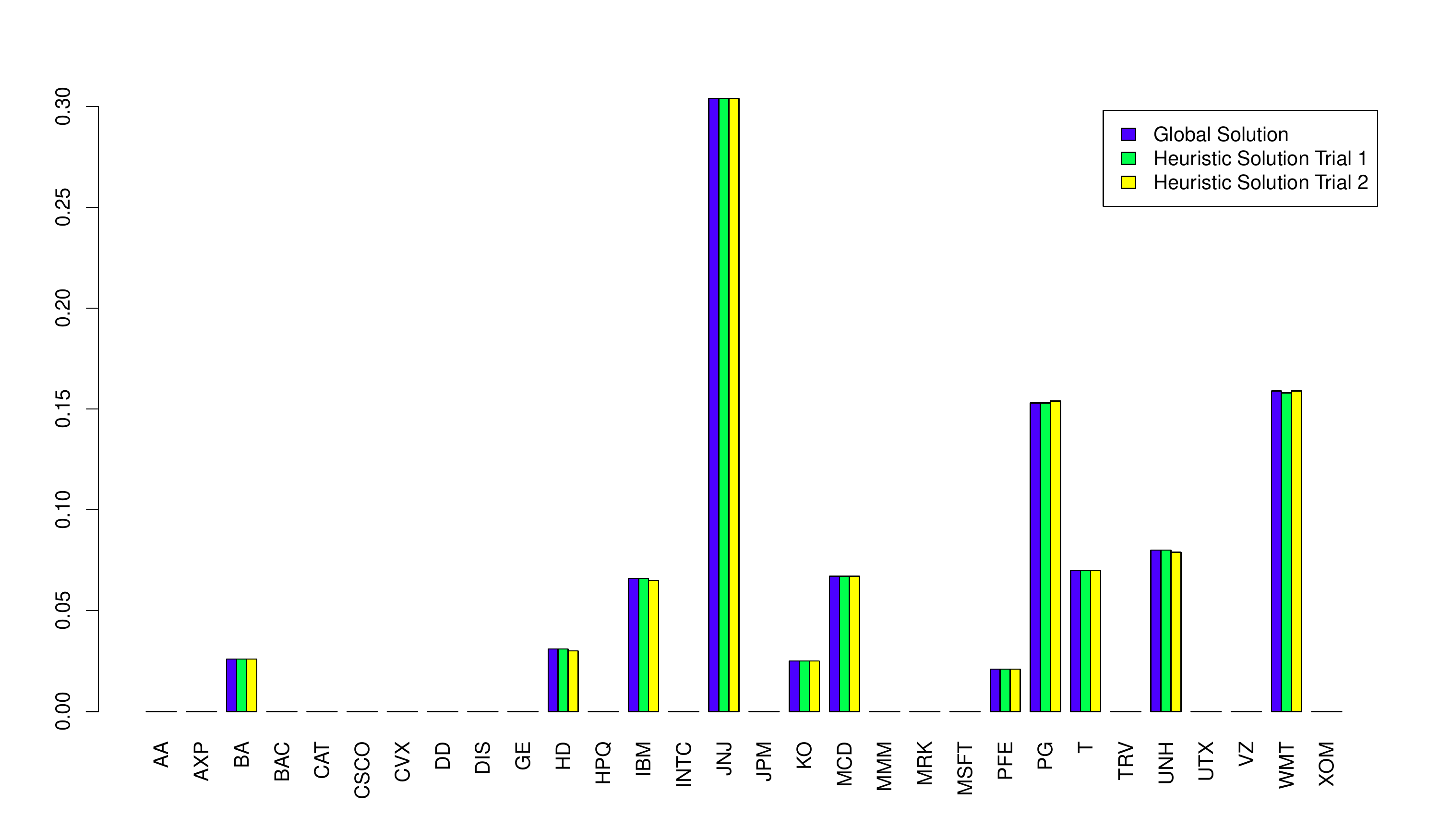}
\caption{Long-only Markowitz-optimal portfolio and solutions of the proposed metaheuristic.}
\label{fig:ex1b}
\end{center}
\end{figure}

Consider the DJIA weekly returns from the year 2012. In Fig.
\ref{fig:ex1a} you can see the global solution computed with a quadratic
solver and two randomly sampled portfolios, which do exhibit the general
structure of a portfolio, but are far off the optimal solution. In Fig.
\ref{fig:ex1b}, you can see the global solution again as well as two
heuristic solutions with $n_1 = 10000$ random samples and a
$\varepsilon$-improvement with $\varepsilon = (0.05, 0.01, 0.001)$.
After (25, 21, 28) as well as (29, 16, 24) iterations, we reach the
global solution. In this simple long-only Markowitz problem, the process
does not need to be iterated.

\subsection{Active-extension CVaR
Portfolios}\label{active-extension-cvar-portfolios}

\begin{figure}
\begin{center}
\includegraphics[scale=0.5]{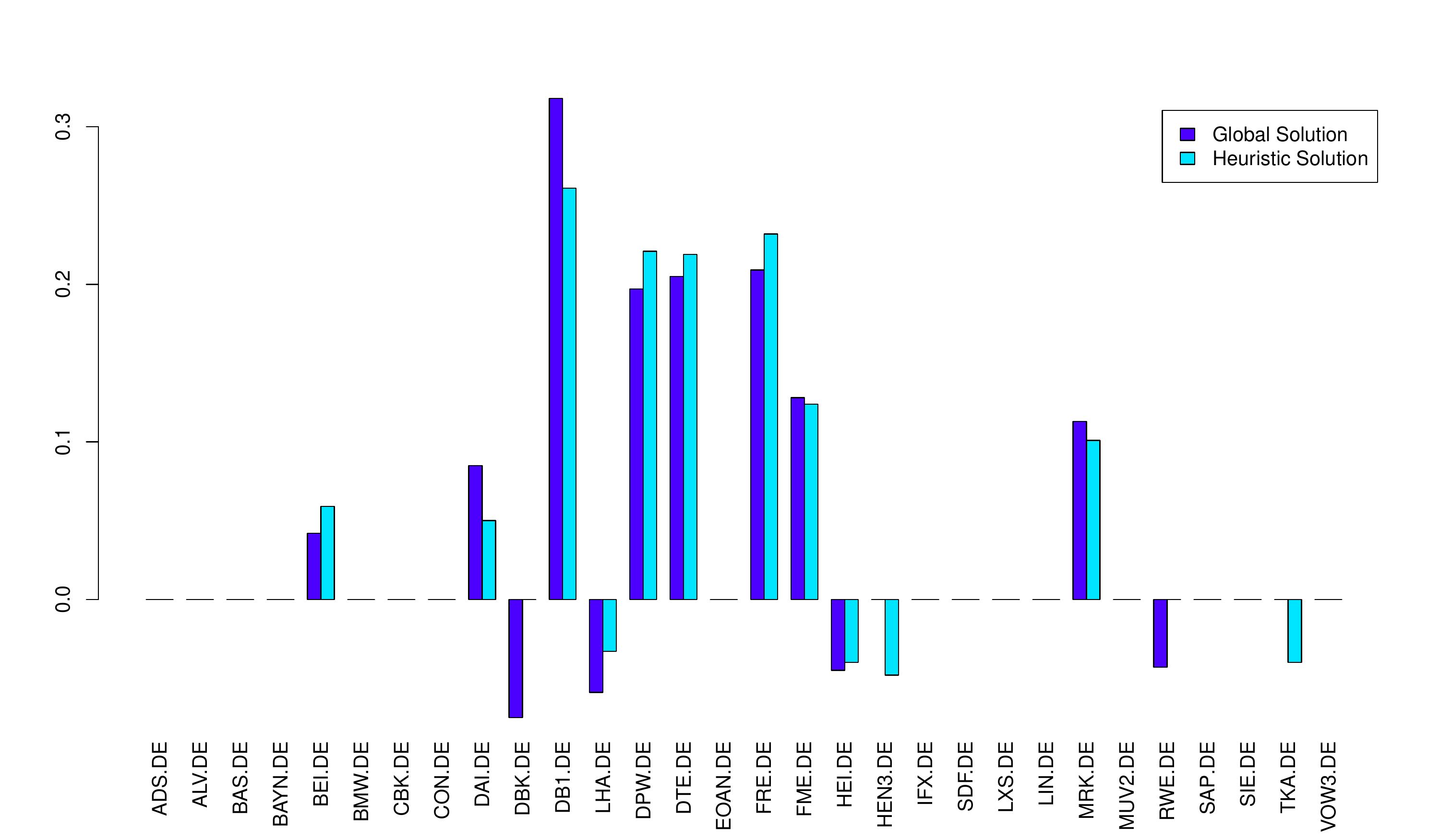}
\caption{Active extension 130/30 CVaR-optimal portfolio.}
\label{fig:ex2}
\end{center}
\end{figure}

In this second example we use DAX 30 weekly returns from the year 2013.
Here we try to compute the CVaR-optimal portfolio, but with active
extension constraints (130/30) and with certain upper and lower bounds
on the assets, i.e.~a maximum of 0.5 long and $-0.1$ short. This turns
out to be more tricky. We start with $10000$ random samples and apply
$\varepsilon$-improvement with $\varepsilon=(0.05,0.02,0.01)$ to the
$n_2 = 10$ best sampled portfolios. The results differ slightly such
that we compute the average and approximate again. The result is shown
in Fig. \ref{fig:ex2}. While the heuristic solution is not completely
similar to the global solution, it is valid from a portfolio managers
point of view, as the optimization result is used as a guideline and
will not be implemented exactly as optimized -- partly because this is
not even possible on real markets.

The portfolio compositions can also be seen in Table \ref{tab:ex2}. In
this table only assets which do exhibit a non-zero allocation in one of
the solutions are shown. In addition, the main risk parameters of the
assets are shown, i.e.~both the standard deviation as well as the
95\%-Value at Risk. Here we can see that e.g.~the assets RWE.DE and
TKA.DE show exactly the same risk structure in these two risk
parameters, such that the misassignment of the heuristic is not too
problematic from an investors point of view.

\begin{table}
\caption{Different portfolio allocations in the 130/30 CVaR case.}
\label{tab:ex2}
\centering
\begin{tabular}{lcccc}
  \hline
 & Std.Dev. & VaR $(95\%)$ & Global Solution & Heuristic Solution \\ 
  \hline
  BEI.DE & 0.02 & $-0.03$ & 0.04 & 0.06 \\ 
  DAI.DE & 0.03 & $-0.04$ & 0.08 & 0.05 \\ 
  DBK.DE & 0.04 & $-0.06$ & \makebox[0pt][r]{$-$}0.07 & 0.00 \\ 
  DB1.DE & 0.03 & $-0.04$ & 0.32 & 0.26 \\ 
  LHA.DE & 0.04 & $-0.05$ & \makebox[0pt][r]{$-$}0.06 & \makebox[0pt][r]{$-$}0.03 \\ 
  DPW.DE & 0.03 & $-0.03$ & 0.20 & 0.22 \\ 
  DTE.DE & 0.03 & $-0.04$ & 0.20 & 0.22 \\ 
  FRE.DE & 0.02 & $-0.03$ & 0.21 & 0.23 \\ 
  FME.DE & 0.03 & $-0.03$ & 0.13 & 0.12 \\ 
  HEI.DE & 0.04 & $-0.06$ & \makebox[0pt][r]{$-$}0.05 & \makebox[0pt][r]{$-$}0.04 \\ 
  HEN3.DE & 0.03 & $-0.03$ & 0.00 & \makebox[0pt][r]{$-$}0.05 \\ 
  MRK.DE & 0.02 & $-0.03$ & 0.11 & 0.10 \\ 
  RWE.DE & 0.04 & $-0.07$ & \makebox[0pt][r]{$-$}0.04 & 0.00 \\ 
  TKA.DE & 0.04 & $-0.07$ & 0.00 & \makebox[0pt][r]{$-$}0.04 \\ 
   \hline
\end{tabular}
\end{table}

\begin{figure}
\begin{center}
\includegraphics[scale=0.5]{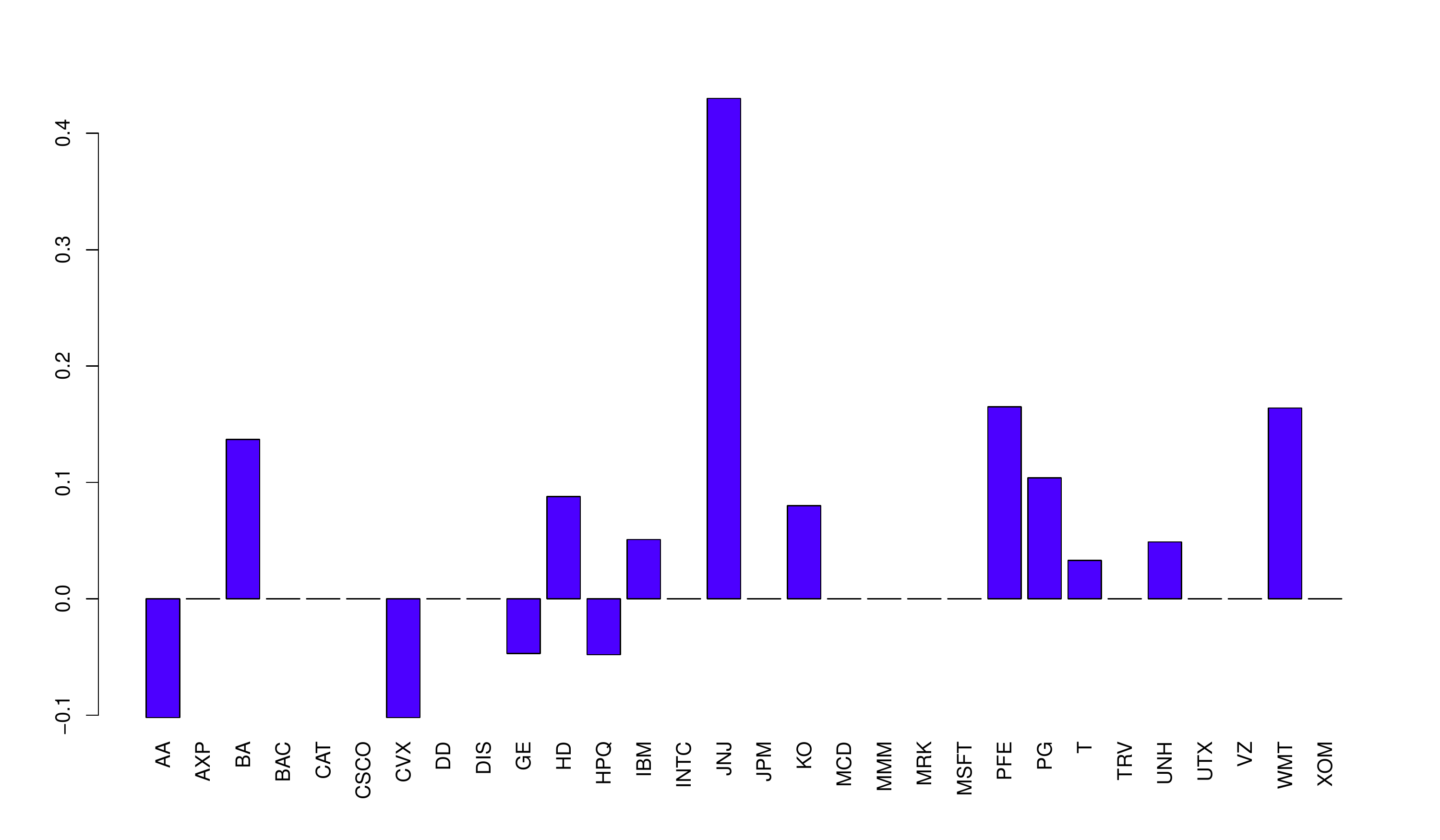}
\caption{Active extension 130/30 Value at Risk-optimal portfolio.}
\label{fig:ex3}
\end{center}
\end{figure}

\subsection{Active-extension VaR
Portfolios}\label{active-extension-var-portfolios}

With the knowledge that our CVaR portfolios are almost spot on from an
investors perspective, we can now apply the same procedure to compute an
active-extension portfolio using non-convex risk-measures. We use data
from the DJIA and apply the same setup as in Section
{[}active-extension-cvar-portfolios{]}, but change the objective
function to minimizing the Value-at-Risk. The result is shown in Fig.
\ref{fig:ex3}.

\section{Conclusion}\label{conclusion}

In this paper we have presented a clever multi-start local search
heuristic for active extension portfolios with convex but also
non-convex risk measures. Despite its simplicity, the metaheuristic
works very well. Future extensions include an automatic tuning of the
set of $\varepsilon$ and further empirical tests, especially with a
larger set of assets.

\bibliographystyle{plainnat}
\bibliography{actext}

\end{document}